\documentstyle[12pt]{article}
\def\v{\begingroup\obeyspaces\u}
\def\u#1{\verb!#1!\endgroup}
\def\IQ{\v{IQ}}

\def\IL{\v{IL}}
\def\ID{\v{ID}}
\def\IV{\v{IV}}
\def\IP{\v{IPROC}}

\def\AL{{\small ALPGEN}}
\def\HD{{\small HDECAY}}
\def\HP{{\small HERWIG++}}
\def\HW{{\small HERWIG}}
\def\PY{{\small PYTHIA}}
\def\IS{{\small ISAJET}}
\def\IW{{\small ISAWIG}}

\def\MS{{\small MSSM}}
\def\MN{{\small MC@NLO}}
\def\SM{{\small SM}}
\def\SY{{\small SUSY}}
\def\ee{$e^+e^-$}
\def\l{\ell}
\def\qbar{\bar q}   
\def\SMH{H^0_{\mbox{\scriptsize SM}}}
\def\yes{\raise0.3em\hbox{$\sqrt{}$}}
\def\no{$\times$}
\begin{document}
\tolerance=100000
\thispagestyle{empty}
\setcounter{page}{0}
\noindent
Cavendish-HEP-02/17\hfill
CERN-TH/2002-270\\
DAMTP-2002-124\hfill
%DCPT/02/16\\
IPPP/02/58\\
KEK--TH-850\hfill
MC-TH-2002-7\\
MPI-PhT 2002-55\hfill
Revised Oct 2005

\begin{center}

{\Large \bf HERWIG 6.5 Release Note}\\[4mm]

{G. Corcella\\[0.4mm]
\it Max-Planck-Institut f\"ur Physik, Werner-Heisenberg-Institut,
Munich\\[0.4mm]
E-mail: \tt{corcella@mppmu.mpg.de}}\\[4mm]

{I.G.\ Knowles\\[0.4mm]
\it Department of Physics and Astronomy, University of Edinburgh\\[0.4mm]
E-mail: \tt{iknowles@supanet.com}}\\[4mm]

{G.\ Marchesini\\[0.4mm]
\it Dipartimento di Fisica, Universit\`a di Milano-Bicocca, and I.N.F.N.,
Sezione di Milano\\[0.4mm]
E-mail: \tt{Giuseppe.Marchesini@mi.infn.it}}\\[4mm]

{S.\ Moretti\\[0.4mm]
\it Theory Division, CERN, and Institute for Particle Physics Phenomenology,
University of Durham\\[0.4mm]
E-mail: \tt{stefano.moretti@cern.ch}}\\[4mm]

{K.\ Odagiri\\[0.4mm]
\it Theory Division, KEK\\[0.4mm]
E-mail: \tt{odagirik@post.kek.jp}}\\[4mm]

{P.\ Richardson\\[0.4mm]
\it Department of Applied Mathematics and Theoretical Physics and\\
Cavendish Laboratory, University of Cambridge\\[0.4mm]
E-mail: \tt{richardn@hep.phy.cam.ac.uk}}\\[4mm]

{M.H.\ Seymour\\[0.4mm]
\it Department of Physics and Astronomy,
University of Manchester\\[0.4mm]
E-mail: \tt{M.Seymour@rl.ac.uk}}\\[4mm]

{B.R.\ Webber\\[0.4mm]
\it Cavendish Laboratory, University of Cambridge\\[0.4mm]
E-mail: \tt{webber@hep.phy.cam.ac.uk}}\\[4mm]

\end{center}

\vspace*{\fill}

\begin{abstract}{\small\noindent
    A new release of the Monte Carlo program \HW\ (version 6.5) is now
    available. The main  new  features  are: support for the Les Houches
    interface to matrix element generators;
    additional \SM\ and \MS\ Higgs processes in lepton collisions;
    additional matrix elements for the spin correlation algorithm;
    a new version of the \IW\ interface; interface to the \MN\ program
    for heavy quark, Higgs and vector boson production in hadron collisions.
    This is planned to be the last major release of Fortran \HW.
    Future developments will be implemented in a new C++ event generator,
    \HP.}

\end{abstract}

\vspace*{\fill}
\newpage
\tableofcontents
\setcounter{page}{1}

\section{Introduction}

The last major public version (6.2) of \HW\ was reported in detail
in \cite{AllHW}.  The new features of version 6.3 and 6.4 are described
in refs.~\cite{Corcella:2001pi} and \cite{Corcella:2001wc}, respectively.
In this note we describe the main modifications and new features
included in the latest public version, 6.5.

Please refer to \cite{AllHW} and to the present paper if
using version 6.5 of the program.
When running \MS\ processes starting from version 6.1, please add 
reference to \cite{Moretti:2002eu}.

The program, together  with other useful files and information,
can be obtained from the following web site:
\small\begin{quote}\tt
            http://hepwww.rl.ac.uk/theory/seymour/herwig/
\end{quote}\normalsize
    This will be mirrored at CERN:
\small\begin{quote}\tt
            http://home.cern.ch/seymour/herwig/
\end{quote}\normalsize

It is anticipated that version 6.5, with minor modifications in the
series 6.5xx, will be the {\em last Fortran version} of \HW.  Substantial
physics improvements will be included in the new C++ event
generator \HP\ \cite{Gieseke:2003hm}.

\section{Les Houches interface}\label{sec:lesh}
    We now  include  support  for the  interface  between  parton  level
    generators and \HW\ using the Les Houches  Accord as  described in
    \cite{Boos:2001cv}. In  general we have  tried to code  the interface in
    such a way that from the user point of  view it behaves in  the same
    way as that already included in \PY\ \cite{Sjostrand:2000wi}.

    The interface  operates in the following  way. If the \IP\ code is
    set  negative  then \HW\  assumes  that the user wants an external
    hard process  using the Les Houches  accord.

    If this  option is  used the  initialization will  call the  routine
    \v{UPINIT} to  initialize  the external hard  process; this  name is the
    same  as that used by \PY. This routine  should set the values of
    the run parameters in the Les Houches run common block.

    After  the  initialization  during the  event  generation  phase the 
    routine  \v{UPEVNT} (again this name is the same as that used by \PY)
    is  called.  This  routine  should  fill the  event  common block as
    described in \cite{Boos:2001cv}.

    Dummy  copies of both these  routines are  supplied with  \HW\ and 
    should be deleted and  replaced if you are using this option. Due to
    the internal  structure of  \HW\ two  new parameters are needed to
    control  the  interface, in  addition  to those  in the  Les Houches
    common block. The logical input variable\footnote{Default values for
    input variables are shown in square brackets.} \v{LHSOFT} [\v{.TRUE}]
    controls the  generation of  the soft underlying event; the default
    is to generate a soft underlying event and so \v{LHSOFT} must be set
    \v{.FALSE.} for lepton-lepton  processes.  The second  variable
    \v{LHGLSF} [\v{.FALSE.}] controls the  treatment of colour
    self-connected  gluons, which  may occur  with some methods of colour
    decomposition. The  default is to kill events with self-connected
    gluons  whereas if \v{LHGLSF} = \v{.TRUE.} an  information-only 
    warning is issued instead.

    We have not included full support for the interface. In  particular
\begin{itemize}
\item It is assumed that all events have two particles of status \v{IDUP}=--1;
\item We do not support  the status codes --2 and --9;
\item The treatment of the code \v{IDUP}=3 is the same as \v{IDUP}=2,
      i.e.\ all intermediate resonances have their masses preserved.
\end{itemize}
    These restrictions are the same as those 
    imposed by \PY. If you have any  problems with the  interface or 
    need the options we have not yet supported please let us know.

    It  should  also be noted  that while  the interface has been tested
    with  Standard  Model  processes, for example  all the  processes in
    \AL\ \cite{Mangano:2002ea}, it is  less well tested
    for \SY\ processes.

\section{New SM and MSSM Higgs processes}
\SM\ and \MS\ Higgs production in association with fermion pairs
in lepton-lepton collisions is now available.
These processes were introduced in \cite{Djouadi:1991tk} and their
phenomenological relevance was discussed in \cite{Djouadi:gp}.
The relevant \IP\ codes are given in table~\ref{tab1}. 

Two new subroutines, \v{HWHIGE} and \v{HWH2HE}, had to be included. 
Fermion masses are retained in the final state
according  to the \HW\ defaults. The same values appear in
the Yukawa couplings. Notice that in the case of charged Higgs boson
production the Cabibbo-Kobayashi-Maskawa mixing matrix has been assumed 
to be diagonal. Furthermore, due to the rather different phase space 
distribution of the final state products, all processes can only be produced 
separately, not collectively. Initial- and final-state radiation (both QED 
and QCD) and beamsstrahlung are included via the usual \HW\ algorithms. 
Finally notice that the use of the \IP\ series 1000 and 1100 for $\l^+\l^-$
processes required some internal modification to \HW, which was 
originally designed to generate leptonic processes only for $\IP< 1000$.
Now we assume an $\l^+\l^-$ process whenever $\IP< 1300$.
These modifications have no implications for the traditional user, but
may affect more knowledgeable ones who have edited previous versions
of the main \HW\ code.
\newpage
\begin{table}[t]
\begin{center}
\begin{tabular}{|c|l|}
\hline
\IP & Process\\
\hline
1000+\ID & $\l^+\l^-\to t\,\bar t\,\SMH$ (\ID\ as in \IP=300+\ID)\\
1110+\IQ & $\l^+\l^-\to q\,\qbar\, h^0$  (\IQ\ as in \IP=100+\IQ)\\
1116+\IL & $\l^+\l^-\to \l^+\l^- h^0$ (\IL=1,2,3 for $e,\mu,\tau$)\\
1120+\IQ & $\l^+\l^-\to q\,\qbar\, H^0$ (\IQ\ as in \IP=100+\IQ)\\
1126+\IL & $\l^+\l^-\to \l^+\l^- H^0$ (\IL=1,2,3 for $e,\mu,\tau$)\\
1130+\IQ & $\l^+\l^-\to q\,\qbar\, A^0$ (\IQ\ as in \IP=100+\IQ)\\
1136+\IL & $\l^+\l^-\to \l^+\l^- A^0$ (\IL=1,2,3 for $e,\mu,\tau$)\\
1140    &  $\l^+\l^-\to d\,\bar u\, H^+ +$ c.c.\\
1141    &  $\l^+\l^-\to s\,\bar c\, H^+ +$ c.c.\\
1142    &  $\l^+\l^-\to b\,\bar t\, H^+ +$ c.c.\\
1143    &  $\l^+\l^-\to e\,\bar\nu_e H^+ +$ c.c.\\
1144    &  $\l^+\l^-\to \mu\,\bar\nu_\mu H^+ +$ c.c.\\
1145    &  $\l^+\l^-\to \tau\,\bar\nu_\tau H^+ +$ c.c.\\
\hline
\end{tabular}
\caption{New \SM\ and \MS\ Higgs processes.}\label{tab1}
\end{center}
\end{table}

\section{Spin correlations in  R-parity violating decays}
    When we included spin correlations in \HW6.4 \cite{Richardson:2001df}
    we did not  include
    either R-parity violating  decays or decays producing  gravitinos in
    the  algorithm. This  led to \HW\  stopping when  such decays were 
    included. This of  course could be  stopped by  switching  the  spin
    correlations  off, i.e.\  \v{SYSPIN}=\v{.FALSE.}.
    We have  now  included the 
    relevant  matrix  elements for R-parity  violating  decays and  hard 
    processes and decays producing gravitinos. At the same  time we have
    made changes so that at both the initialization and event generation
    stages many of the terminal warnings  which were caused  by the code
    not having the correct matrix elements are now information-only
    warnings. If you still  get terminal  error messages from any of the
    spin correlation  routines please let us know.

\section{Other interfaces}
\subsection{ISAWIG}
    To  coincide with  the release of  \HW\ 6.5 we have  produced a new
    version  of the \IW\ interface to  \IS\ for the  calculation of
    \MS\ spectra and  decay rates.  Due to stability  problems with the 
    Oxford  web-server  we have moved the web-page to  Cambridge. The new
    address is 
\small\begin{quote}
    \v{http://www.hep.phy.cam.ac.uk/}$\sim$\v{richardn/HERWIG/ISAWIG/}
\end{quote}\normalsize
    At  the moment  the old page will redirect  users here but we cannot
    be sure how long this will continue.

    Since the original version of \IW\ there have been a number of new
    versions of \IS, often with changes in the  sizes of  the common
    blocks from which we extract the information we need. This has
    necessitated periodical updating of the code to run with the most
    recent version  of  \IS, with the result that the code could
    no longer be run with older versions. However, the Snowmass points and
    slopes \cite{Allanach:2002nj} are  defined  with  \IS\ version
    7.58 and so we  can no  longer
    continue  in  this way and  still be able to generate  these  points.
    Therefore, starting  with  the  new  \IW\ version  1.2, we are using C
    preprocessing so that users can define the version of \IS\ they
    are using at  compile time.  When compiling the main \IW\ code and 
    the modified {\small SUGRUN} and {\small SSRUN}
    programs the following compiler options should 
    be specified:
\begin{itemize}
\item \v{-DISAJET758}  to use \IS7.58, 
\item \v{-DISAJET763}  to use \IS7.63.
\end{itemize}
    The default at the moment is to compile code to run with \IS7.64.
    In  the future  this will change  so that the most recent version of 
    \IS\ becomes the default but we will continue to support the older
    versions.

\subsection{HDECAY}
    There has also been a new release of the \HD\ package \cite{Djouadi:1997yw},
    version 3.
    In order to support this  version, and the  previous version 2.0, we 
    are also using C  preprocessing  to control  the version  of  \HD\
    used. In order to achieve  this the \HD\  interface code has  been
    merged with the main \IW\ program. The following options should be
    used when compiling \IW\ if you are using \HD:
\begin{itemize}
\item \v{-DHDECAY2} if you are using version 2 of \HD
\item \v{-DHDECAY3} if you are using version 3 of \HD.
\end{itemize}
    As  before  the default  is to  compile a  dummy routine. If you are
    using \HD\ you must  link with a version of the \HD\  code which 
    has the main \HD\ program removed.

\subsection{MC@NLO}
  The program \MN\ \cite{Frixione:2002ik,Frixione:2003ei,Frixione:2005gz}
generates events with
\begin{itemize}
\item  exclusive rates and distributions accurate to next-to-leading order
    when expanded in $\alpha_s$;
\item multiple  soft/collinear  parton emission generated by \HW\ parton
    showering;
\item hadronization according to the \HW\ cluster model.
\end{itemize}

%%%%%%%%%%%%%%%%%%%%%%%%%%%%%%%%%%%%%%%%%%%%%%%%%%%%%%%%%%%%%%%%%%%%%%%%
\begin{table}[htb]
\begin{center}
\begin{tabular}{|c|c|c|c|c|l|}\hline
\IP & \IV & \IL$_1$ & \IL$_2$ & 
 Spin & Process \\\hline
 --1350--\IL & & & &\yes &
 $H_1 H_2\to (Z/\gamma^*\to) l_{\rm IL}\bar{l}_{\rm IL}+X$\\\hline
 --1360--\IL & & & &\yes &
 $H_1 H_2\to (Z\to) l_{\rm IL}\bar{l}_{\rm IL}+X$\\\hline
 --1370--\IL & & & &\yes &
 $H_1 H_2\to (\gamma^*\to) l_{\rm IL}\bar{l}_{\rm IL}+X$\\\hline
 --1460--\IL & & & &\yes &
 $H_1 H_2\to (W^+\to) l_{\rm IL}^+\nu_{\rm IL}+X$\\\hline
 --1470--\IL & & & &\yes &
 $H_1 H_2\to (W^-\to) l_{\rm IL}^-\bar{\nu}_{\rm IL}+X$\\\hline
 --1396 & & & &\no &
 $H_1 H_2\to \gamma^*(\to \sum_i f_i\bar{f}_i)+X$\\\hline
 --1397 & & & &\no &
 $H_1 H_2\to Z^0+X$\\\hline
 --1497 & & & &\no &
 $H_1 H_2\to W^+ +X$\\\hline
 --1498 & & & &\no &
 $H_1 H_2\to W^- +X$\\\hline
 --1600--\ID & & & & &
 $H_1 H_2\to H^0+X$\\\hline
 --1705 & & & & &
 $H_1 H_2\to b\bar{b}+X$\\\hline
 --1706 & & & & \no &
 $H_1 H_2\to t\bar{t}+X$\\\hline
 --2600--\ID & 1 & 7 & &\no &
 $H_1 H_2\to H^0 W^+ +X$\\\hline
 --2600--\ID & 1 & $i$ & &\yes &
 $H_1 H_2\to H^0 (W^+\to)l_i^+\nu_i +X$\\\hline
 --2600--\ID & --1 & 7 & &\no &
 $H_1 H_2\to H^0 W^- +X$\\\hline
 --2600--\ID & --1 & $i$ & &\yes &
 $H_1 H_2\to H^0 (W^-\to)l_i^-\bar{\nu}_i +X$\\\hline
 --2700--\ID & 0 & 7 & &\no &
 $H_1 H_2\to H^0 Z +X$\\\hline
 --2700--\ID & 0 & $i$ & &\yes &
 $H_1 H_2\to H^0 (Z\to)l_i\bar{l}_i +X$\\\hline
 --2850 & & 7 & 7 & \no &
 $H_1 H_2\to W^+W^-+X$\\\hline
 --2850 & & $i$ & $j$ & \yes &
 $H_1 H_2\to (W^+\to)l_i^+\nu_i (W^-\to)l_j^-\bar{\nu}_j +X$\\\hline
 --2860 & & 7 & 7 & \no &
 $H_1 H_2\to Z^0Z^0+X$\\\hline
 --2870 & & 7 & 7 & \no &
 $H_1 H_2\to W^+Z^0+X$\\\hline
 --2880 & & 7 & 7 & \no &
 $H_1 H_2\to W^-Z^0+X$\\\hline
\end{tabular}
\end{center}
\caption{\label{tab:proc} 
Processes implemented in \MN\ version 3.1.}
\end{table}
%%%%%%%%%%%%%%%%%%%%%%%%%%%%%%%%%%%%%%%%%%%%%%%%%%%%%%%%%%%%%%%%%%%%%%%%

The processes implemented in the current version (3.1) of \MN\ are
listed in Table~\ref{tab:proc}. Here $H_{1,2}$ represent incoming
hadrons, $H^0$ denotes the Standard Model
Higgs boson and the value of \ID\ controls its decay, as described
in the \HW\ manual. The values of \IV, \IL,
\IL$_1$, and \IL$_2$ control the identities of vector
bosons and leptons: for details see the \MN\ manual~\cite{Frixione:2005gz}.
\IP--10000 generates the same processes as \IP, but eliminates the underlying
event. A void entry indicates that the corresponding variable is unused. The
`Spin' column indicates whether spin correlations in vector boson or top
decays are included (\yes), neglected (\no) or absent (void entry). Spin
correlations in Higgs decays are included by \HW\ (e.g. in $H^0\to W^+W^-\to
l^+\nu l^-\bar{\nu}$).

Parton configurations must first be generated by the modified NLO
program provided in the \MN\ package.  These are stored in a file which is
then read into \HW\ via the Les Houches interface.
For further details and updates see the \MN\ web page:
\small\begin{quote}\tt    
            http://www.hep.phy.cam.ac.uk/theory/webber/MCatNLO/
\end{quote}\normalsize

\section{Miscellaneous changes and corrections}
\begin{itemize}
\item
    A new logical input variable, \v{PRESPL} [\v{.TRUE.}],  has  been
    introduced to control whether  the  longitudinal  momentum 
    (\v{PRESPL} = \v{.TRUE.}),  or  rapidity (\v{PRESPL} = \v{.FALSE.}),
    of the hard process centre-of-mass is preserved in
    hadron collisions after initial-state parton showering.   At present
    the only function of this variable is to allow users to study the
    effects of momentum reshuffling,  which is necessary after showering
    to compensate for jet masses. In future, it is anticipated that setting
    \v{PRESPL}=\v{.FALSE.} will simplify the treatment of other processes
    in \MN.

\item A bug has been  fixed  in the  backward  evolution in the  logical
      structure structure of an \v{IF} statement controlling the branching.
      This was  preventing the forced  branching  to valence partons and
      caused many  of the \v{HWSBRN}=104  warnings.  Users  should  find the
      number of such warnings is significantly reduced.

\item When polarization  effects in lepton  collisions  were added a bug 
      was introduced in squark pair production which has now been fixed.

\item A bug in the matrix element for $q\,\qbar\to g\,\SMH$ (\IP=2300
      etc.) has been corrected.

\item The effects of off-shell $WW$ and $ZZ$ decays of the SM Higgs boson
      are now included in the generated weights of $2\to 3$ processes.
\end{itemize}
We list here the further changes made between the release of version
6.500 and the most recent version, 6.510 (October 31st 2005).

\pagebreak[3]\noindent
In {\bf 6.503}, only bug fixes and minor changes were made:
\begin{itemize}
\item Alignment of initial state radiation cones
(affects HWBJCO)
\item Some things needed by MC@NLO
(affects HWBJCO)
\item Bug fix in initial-state spin correlations
(affects HWBSPA)
\item Bug fix in finding gauge boson pairs
(affects HWDBOS)
\item Bug fix in heavy object decay correlations
(affects HWDHO2)
\item Les Houches underlying event 
(affects HWHGUP)
\item Bug fix in Les Houches interface for $2\to1$ processes
(affects HWHGUP)
\item Bug fix: top lifetime
(affects HWUDAT)
\item Bug fix for tops with Les Houches
(affects HWHGUP)
\item Bug fix in spinor routines to avoid divide by zero
(affects HWH2F1, HWH2F2, HWH2F3)
\item Bug fix in error severity for negative energy underlying events
(affects HWMULT)
\item Bug fix for Les Houches Higgs decay
(affects HWDHIG)
\end{itemize}

\pagebreak[3]\noindent
In {\bf 6.504}, a couple of the changes were slightly more significant:
\begin{itemize}
\item The use of the running quark mass for $q\bar q\to$ Higgs, this 
is to avoid the unphysically large contribution to cross section $u\bar u\to$
Higgs was giving using the constituent quark masses. (New routine HWURQM
and affects HWHIGS).  Note that this is not done for any other Higgs
production or decay processes.
\item Changes to allow the forcing of gauge boson decays in 
   processes using the Les Houches interface (affects HWDBOS)
\item A new flag ITOPRD to allow the use of PHOTOS to generate QED radiation
   in top quark pair production and decay. The default ITOPRD=0 is not to
   use PHOTOS whereas ITOPRD=1 will use PHOTOS. It should be noted that 
   there may be a problem with double counting if the default HERWIG 
   photon radiation in the parton shower is also switched on, however the 
   new option will give radition for the leptons in the decay of the W 
   which the HERWIG treatment does not.
(New routines HWPHTP and HWPHTT and affects HWBCON, HWBTOP, HWDHO4, HWDTAU
and HWUINC).
\end{itemize}

\pagebreak[3]\noindent
In {\bf 6.505}, one new feature was added and several minor improvements
and bug fixes were made:
\begin{itemize}
\item New feature:
\begin{itemize}\item A major update of the Jimmy generator for multiple parton scattering
  has been made.  Although this is still a separate package (available
  from the Jimmy web page)
  its incorporation into
  HERWIG is now much smoother - no HERWIG routines need to be modified
  or replaced.  Moreover, it has been modified to run correctly in
  'underlying event' mode, i.e. it can attach additional scatters to a
  high-pt scattering event or to events of other types, rather than
  always running in 'minimum bias' mode as before
\end{itemize}
\item Improvements:
\begin{itemize}\item Value of MODMAX (size of MODBOS) increased to 50.  Note that since
  MODBOS is the last member of its common block, this does not move the
  position of any other variables in the include file
\item Calculation of minimum invariant mass needed for a given hard process
  separated off from HWEGAM to a new routine HWEGAS
\item Several new features added to HWHSCT and new routine HWHSCU added to
  implement the underlying event mode of Jimmy
\end{itemize}
\item Bug fixes:
\begin{itemize}\item Several modifications (in HWCFOR, HWCHAD and HWURES) to improve the
  calculation of the threshold for partonic decays of $b$ baryons and $B_c$
  mesons
\item Hard scale variable EMSCA was not set correctly in gauge boson pair
  production (HWHGBS and HWHGBF)
\item In many places in the code, in several recently added processes, tests
  on IPROC were not performed correctly, leading to different behaviour
  in the same hard process, depending on whether the underlying event
  was switched on (IPROC${}<10000$) or off (IPROC${}>10000$)
\item Probability of backwards evolution to an 'anomalous' photon was not
  correctly calculated.  New probability is somewhat larger at large
  x-gamma, but similar at small x-gamma (HWSFBR)
\item Protection against corruption of photon production cross sections due
  to infinitesimal parton distribution functions (HWHPHO and HWHPH2)
\item Corrected angular distributions in $gg\to tbH$ (HWHIGQ)
\item Corrected cross section in Higgs+jet production (HWHIGA)
\item Added protection against clusters of unknown flavour combinations
  (HWCHAD)
\end{itemize}\end{itemize}

\pagebreak[3]\noindent
In {\bf 6.506}, one new feature was added and several minor improvements
and bug fixes were made:
\begin{itemize}
\item New feature:
\begin{itemize}\item Two new parameters, PDFX0 and PDFPOW, have been provided to control
   the probing of parton distribution functions at extremely small x
   values, potentially below where they are valid.  For values of x
   below PDFX0, $x f(x,Q^2)$ is replaced by $PDFX0 f(PDFX0,Q^2) (x/PDFX0)^{PDFPOW}$.
   The default value of PDFX0 (D=0) means that the pdfs are unmodified
   and that the value of PDFPOW (D=0) is irrelevant.  For 'valence-like'
   distributions at small x, set PDFX0=1d-5 for example, and PDFPOW=0.
   This feature is mainly needed in conjunction with Jimmy since, for
   PTJIM=2GeV at the LHC for example, x values down to $2\times10^{-8}$ are
   probed, whereas most pdf sets are not considered reliable below x
   values of about $10^{-5}$
\end{itemize}
\item Improvements:
\begin{itemize}\item Spin correlations in the decay of the W/Z in W/Z+Higgs events are now
   included
\item In several gauge boson production processes, hard coded limits on the
   mass distribution have been replaced by GAMMAX [D=10] widths
\item Generation of Breit-Wigner distribution is changed from $m$ to $m^2$
\item Infrared cutoffs on large x values implemented in Drell-Yan matrix
   element correction routine HWBDYP.  Makes a negligible difference
\item Running electromagnetic coupling used in W+jet routine
\item Branching fraction replaced by running partial width divided by m
   times nominal total width in W production routine HWHWPR.  This is
   only different in the rare events in which the chosen W mass is above
   the top threshold
\item More accurate calculation of hadron remnant mass used in HWSBRN
\item PDFSET is only called if the pdf set has changed since the last call,
   saving cpu time
\end{itemize}
\item Bug fixes:
\begin{itemize}\item Kinematics were not calculated correctly in Drell-Yan matrix element
   correction routine HWBDYP.  Has a tiny effect on W/Z kinematics, but
   more significant on the distributions of their decay products, giving
   too hard a lepton pt distribution at very high pt for example
\item In several Higgs and/or gauge boson production processes, including
   W+jet, W/Z+H and H+jet, Breit-Wigner was generated twice, leading to
   a too wide mass distribution by a factor of SQRT(2)
\item Mass distribution of W/Z in W/Z+H and W/Z+jet production were
   incorrect by a factor $m^2/m_0^2$, where $m_0$ is the nominal mass
\item Subprocess $bg\to Wt$ was previously included in W+jet routine in the
   massless quark approximation.  This subprocess has been switched off
\item Off-shell hadron remnants sometimes led to momentum non-conservation
   in secondary scatters generated by Jimmy.  Fixed by shuffling small
   amount of momentum between the two remnants in HWHREM
\end{itemize}\end{itemize}

\pagebreak[3]\noindent
In {\bf 6.507}, only bug fixes were made:
\begin{itemize}
\item Bug fixes:
\begin{itemize}\item Space-time production position of hadrons from the underlying event
   was previously non-sensical.  Now the cluster position is chosen
   according to a Gaussian distribution in its rest-frame and the
   hadron positions correctly take account of the position of the
   primary interaction point.  Implemented in HWMEVT
\item Space-time production position of leptons and photons from gauge
   and Higgs boson decays was previously set to the origin.  Now
   calculated correctly taking account of the position of the primary
   interaction point.  Implemented in HWBJCO, HWCFOR, HWDBOS and
   HWDHIG
\item A bug has been found in the interface to Jimmy that led to errors
   (stable quarks in the final state and/or divide-by-zero crashes)
   when adding multiple scatters to Drell-Yan-type processes.  Fixed
   in HWHSCT
\item MRST pdfs did not previously freeze at QSPAC in the ISPAC.GT.0
   options, as they should.  Fixed in HWSFUN
\item Printed version number was incorrect in several places in the LaTeX
   and html output formats.  Fixed in HWUDPR and HWUEPR
\item A minor improvement to the formatting of the printed event record
   for minimum bias events, implemented in HWUEPR
\end{itemize}\end{itemize}

\pagebreak[3]\noindent
In {\bf 6.510}, only bug fixes and improvements to the machine/compiler
dependence were made:
\begin{itemize}
\item Bug fixes:
\begin{itemize}\item A bug has been found in the kinematic reconstruction of parton
   showers, that had particularly severe consequences for top quark
   decay.  This step involves reshuffling a small amount of momentum
   between jets in order to restore overall momentum conservation.  It
   is not Lorentz invariant, so the jet momenta must be boosted to and
   from the frame in which it is performed.  In versions 6.504 to 6.507,
   these two boosts were performed via different intermediate frames,
   and therefore a (Thomas) rotation was induced.  This lead to a
   significant shift in the direction of the b jet, even in events in
   which there was no radiation from it.\\
   Note that this fix has already been circulated to several experimental
   collaborations, who were asked to call the fixed version 6.508.
\item A bug in the logic of the azimuthal correlations between two
   back-to-back jets was present in versions 6.503 to 6.507, leading to
   a slight (a few parts per mille) asymmetry in the azimuthal
   distribution of produced hadrons in hadron collisions.
\item A minor bug was fixed in tau decays using TAUOLA, which had lead to
   Lorentz-noninvariance of some spin effects.
\item A variable, TMPRN, was used, but not declared, as an array in HWHSCT,
   leading to crashes or unpredictable results when using Jimmy on some
   machines.
\item Our particle numbering distinguishes Standard Model and BSM Higgs
   bosons, while the pdg use the same number for the SM and lightest
   CP even BSM Higgs.  Our numbering is now only used internally and is
   converted to the pdg convention at the end of event processing.
\end{itemize}
\item Changes made to reduce machine/compiler dependence:
\begin{itemize}\item Multiple-ENTRY routines have been replaced by independent routines.
\item Alternate return points have been removed.  Note that in particular,
   this affects HWWARN, so users who use HWWARN should change their
   call accordingly:
\begin{verbatim}
      CALL HWWARN('HWANAL',ICODE,*999)
\end{verbatim}
   should become
\begin{verbatim}
      CALL HWWARN('HWANAL',ICODE)
      IF (ICODE.LT.0.OR.(ICODE.GE.50.AND.ICODE.LT.200)) GOTO 999
\end{verbatim}
\item An arithmetic IF statement has been removed.
\item All variables in DATA statements have also been SAVEd.
\item Line numbers have been removed from END statements.
\end{itemize}
Several of these are necessary to compile under gcc4.0.  Note that the
first point necessitated moving the random number seeds from a SAVEd
local variable to a COMMON block.  We had some concerns that this may
slow the program, but in fact on several compilers and machines we find
that the new version is actually faster.  Nevertheless, if you encounter
significantly slower performance with the new version, please inform us,
with details of the machine, compiler and compiler options.\\
We are grateful to Mikhail Kirsanov and the GENSER group of the LCG
project at CERN for their help with some of these changes.
\end{itemize}

\section*{Acknowledgements}
Since this is expected to be the last (Fortran) \HW\ release note, we would
like to take this opportunity to thank again all those colleagues and users
who have contributed to the development of the program, whether by providing
code, suggesting improvements or reporting problems.  Thanks also to those
who worked so hard to establish the Les Houches accord, which should make
it possible to expand the application of Fortran \HW\ to new processes
without (much) further intervention by the authors.  Meanwhile, many of us
will be transferring our main efforts to \HP, in preparation for the new
era of object-oriented event generation.

\end{document}